\documentstyle[12pt]{article}

\def\void{}
\def\labelmark{}

\newenvironment{formula}[1]{\def\labelname{#1}
\ifx\void\labelname\def\junk{\begin{displaymath}}
\else\def\junk{\begin{equation}\label{\labelname}}\fi\junk}%
{\ifx\void\labelname\def\junk{\end{displaymath}}
\else\def\junk{\end{equation}}\fi\junk\labelmark\def\labelname{}}

{\ifx\void\labelname\def\junk{\end{array}\end{displaymath}}
\else\def\junk{\end{array}\right.\end{equation}}
\fi\junk\labelmark\def\labelname{}\def\junk{}
}

\newcommand{\beq}{\begin{formula}}
\newcommand{\eeq}{\end{formula}}
\newcommand{\beqv}{\begin{formula}{}}

\newcommand{\rf}[1]{(\ref{#1})}

\newcommand{\bea}{\begin{eqnarray}}
\newcommand{\eea}{\end{eqnarray}}
\newcommand{\beas}{\begin{eqnarray*}}
\newcommand{\eeas}{\end{eqnarray*}}
\newcommand{\beqs}{\begin{displaymath}}
\newcommand{\eeqs}{\end{displaymath}}

\newcommand{\br}{\langle}
\newcommand{\kt}{\rangle}
\newcommand{\nn}{\nonumber \\}
\newcommand{\cH}{{\cal H}}
\newcommand{\cG}{{\cal G}}
\newcommand{\cE}{{\cal E}}

\newcommand{\cL}{{\cal L}}

\newcommand{\cM}{{\cal M}}

\newcommand{\Tr}{{\rm Tr}\;}
\newcommand{\ben}{\begin{equation}}
\newcommand{\een}{\end{equation}}

\newcommand{\bdm}{\begin{displaymath}}
\newcommand{\edm}{\end{displaymath}}

\newcommand{\pa}{\partial}
\newcommand{\bbZ}{{\bf Z}}
\newcommand{\bR}{{\bf R}}

\begin{document}
 \topmargin 0pt
 \oddsidemargin 5mm
 \headheight 0pt
 \topskip 0mm

 \addtolength{\baselineskip}{0.4\baselineskip}

 \pagestyle{empty}

 \vspace{1cm}

\hfill RH-05-99

\vspace{2cm}

\begin{center}

{\Large \bf The Spectrum of a }

\medskip

{\Large\bf Transfer Matrix for Loops}

%\vspace{1 truecm}

%\centerline{}

\vspace{1.2 truecm}

%\vspace{1.5cm}

%DRAFT

 \vspace{1.2 truecm}

  Thordur Jonsson\footnote{e-mail: thjons@raunvis.hi.is} 

 University of Iceland

 Dunhaga 3, 107 Reykjavik

 Iceland

 \vspace{1.5 truecm}

 George K. Savvidy\footnote{e-mail: savvidy@argos.nuclear.demokritos.gr}

 NRCPS ``Demokritos''

 Ag. Paraskevi, 15310 Athens

 Greece

 \vspace{1.2 truecm}

 \end{center}

 \noindent
 {\bf Abstract.} We study the spectral properties of the transfer matrix for
 a gonihedric random surface model on a three-dimensional lattice.  
The transfer
 matrix is indexed by
 generalized loops in a natural fashion and is invariant under 
 a group of motions in loop-space.
 The eigenvalues of the transfer matrix can be evaluated exactly in terms of 
 the partition function, the internal energy 
and the correlation functions of the two-dimensional 
 Ising model and the corresponding eigenfunctions are explicit 
functions on loop-space.

 \vfill

 \newpage
 \pagestyle{plain}

 \section{Introduction}
 The triviality of simple random surface models on the lattice, due to 
 the dominance of branched polymers \cite{durhuus}, and the 
 non-scaling of the string tension in analogous Euclidean invariant models 
 based on dynamical triangulations \cite{ambjorn} 
 made it imperative to study models 
 with an action depending on the extrinsic curvature of surfaces.
 For a review of work in this area up to 1997, see \cite{book}.

 One of the models with extrinsic curvature dependent action 
 is the so-called gonihedric random surface model
 introduced in \cite{ambartzumian} and further studied in 
 \cite{savvidy1,savvidy2,savvidy3}.  This model
 is unstable in the simplest cases \cite{durhuus2}.
 However, there is a lattice 
 discretization of the model \cite{savvidy4,savvidy5} which is not
 plagued by stability problems and is more amenable to analytical study, see
 \cite{savvidy6}-\cite{pietig2}.  
 Another lattice model with similar
 properties was introduced and studied in \cite{durhuus3}.  

 In \cite{jonsson}  a simplified version of the three-dimensional 
 gonihedric lattice model was studied 
 and the two largest eigenvalues of the transfer matrix 
were determined.  In this 
 paper we finish the calculation of the spectrum of the transfer matrix
 for the model and 
 find all the eigenfunctions explicitly.

 \section{Transfer matrix for loops} 

 Let $T_3$ denote a sublattice of $\bbZ ^3$ of size $M\times M\times N$ with
 periodic boundary conditions.  We shall think of the third coordinate
 direction where the lattice has extension 
 $N$ as the `time' direction.  The
 configuration space of the system we wish to study is a collection of 
 subsets of the
 plaquettes in $T_3$, denoted $\cM$,  which we refer to as singular surfaces or
 simply as {\em surfaces}.
 A collection of plaquettes $M$ belongs to $\cM$ if and only if any link
 in a plaquette in $M$ belongs to an even number of plaquettes in $M$. 
 This means that the surfaces in $\cM$ do not overlap themselves but they
 can intersect themselves at right angles 
 and they are closed.  Note that the surfaces
 need not be connected.  It is not hard to see
 that the conguration space $\cM$ is identical to the configuration space
 of the three-dimensional Ising model on the periodic lattice $T_3$.
The surfaces are in one to one correspondence with phase boundaries.

 We say that a link $\ell$ in a surface $M\in\cM$ is an {\em edge} 
 if $\ell$ is
 contained in exactly two plaquettes that meet at a right angle.  We denote the
 collection of all edges in $M$ by $\cE (M)$.  The action $S(M)$ of the surface
 $M$ is defined as the total number of edges in $M$
 \beq{1}
 S(M)=\# \cE (M).
 \eeq
 Note that no action is associated with those links where the surfaces 
 crosses itself.
 The partition function is given by
 \beq{2}
 Z(\beta )=\sum _{M\in\cM}e^{-\beta S(M)}.
 \eeq
We define an {\it intermediate plane} 
in $\bR ^3$  to be a plane that lies parallel to but 
 in between planes where one of the coordinates takes an integer value. 
 If we slice a surface $M$ (regarded as a subset of $\bR ^3$) 
 by an intermediate plane
 we obtain a collection of links which we shall call generalized
 loops or simply {\em loops}.  The collection of all possible 
loops that can arise in this way will be denoted
 $\Pi$.  A collection of links $P$ in an intermediate plane belongs to
 $\Pi$ exactly when each vertex in $P$ belongs to an even number of
 links in $P$.  We note that the loops $\Pi$ coincide
 with the phase boundaries that arise in the two-dimensional Ising model.
 It will not come as a surprise that this close connection to the Ising
 model will make the present model exactly soluble in a very strong sense.

 For $P\in\Pi$ we let $|P|$ denote the number of links in $P$.  Note that
 if
 $P_1,P_2\in\Pi$, then the symmetric difference
 \beq{3}
( P_1\cup P_2)\setminus (P_1\cap P_2)\equiv P_1\triangle P_2\in\Pi
 \eeq
 is again a loop in $\Pi$.
 It is easy to check that 
 \beq{3x}
 |P_1\triangle P_2|=|P_1|+|P_2|-2|P_1\cap P_2|
 \eeq
 and $d(P_1,P_2)=|P_1\triangle P_2|$ is a metric on $\Pi$.

 Let $k(P)$ denote the number of corners of $P$, i.e.\ the number of
 vertices of order 2 whose adjacent links are at right angles.  
 Let us denote by $\cH$ the space of real-valued functions on $\Pi$.  Let
 $K_\beta$ be the operator on $\cH$ whose matrix (kernel) is given by
\cite{savvidy6}
 \beq{4}
 K_\beta (P_1,P_2)=\exp \left(-\beta (k(P_1)+k(P_2)+2|P_1\triangle
 P_2|)\right). 
 \eeq
 With this notation the partition function of the model can be expressed as
 \beq{5}
 Z_N(\beta )=\Tr K_\beta ^N
 \eeq
 by slicing surfaces in intermediate planes orthogonal to the time direction.
 This partition function can be evaluated if we ignore the interesection
 term $|P_1 \cap P_2|$ in the exponent of $K_\beta$ 
or if we replace $k(P_1)+k(P_2)$ by
 $k(P_1\triangle P_2)$ \cite{savvidy6}.  
 In this approximation the 
  model becomes a stack of
 noninteracting two-dimensional models.  The
 critical beahaviour of the model is identical to that of the
 two-dimensional Ising model and the free energy can be computed exactly
 by standard methods \cite{baxter}.

 Here we will make an approximation different from the one of
 \cite{savvidy6}.  We drop the curvature terms in the action 
 and study the simplified transfer matrix
 \beq{6}
 K_\beta (P_1,P_2)=\exp\left( -2\beta (|P_1\triangle P_2|-M^2)\right),
 \eeq
 where we have inserted a normalization factor $M^2$ in the exponent for later
 convenience.  
 Dropping the curvature term is the same as neglecting the
 action associated with edges in the time-direction.
 We note the mathematical 
analogy of the transfer matrix \rf{6} with the transition function
 $\exp (-\beta |x-y|)$, $x,y\in\bR ^d$, for random walk in $\bR ^d$.
 This makes it reasonable to regard $K_\beta$ as describing the 
 diffusion of loops.

 With abuse of notation we denote the new transfer matrix \rf{6} by
 the same symbol as
 the old one \rf{4}.  
This should not cause any confusion since we stick to the 
 notation of Eq.\ \rf{6} in the sequel.

 \section{Eigenvalues of the transfer matrix}

 In this section we solve the eigenvalue problem
 \beq{6x}
 \sum_{Q\in \Pi}K_\beta (P,Q)\psi_i(Q)=\Lambda_i(\beta )\psi_i(P),
 \eeq
 $\psi_i\in\cH$.
 We prove that only the eigenvalues depend on $\beta$, 
 not the eigenfunctions.  The eigenvalues can all be expressed in terms of
 the partition function, the internal energy 
 and the spin correlation functions  of the two-dimensional Ising model.  
The eigenfunctions are explicit functions
 on $\Pi$ which can be normalized to take only the values $1$ and $-1$.

 Note first that
 \bea
 \sum_QK_\beta (P,Q) &=& \sum _Qe^{-2\beta (|Q|-M^2)}\nn
                     &=& \Lambda _0(\beta),\label{7y}
 \eea
 where we have in the first step shifted the summation variable from $Q$
 to $P\triangle Q$ (permissible since the mapping $P\mapsto P\triangle Q$
 is bijective) and $\Lambda_0 (\beta)$ is the partition function of a
 two-dimensional Ising model on a periodic $M\times M$ lattice.
 This lattice will be denoted $T_2$.  
 The loops $Q$ are the
 phase boundaries of the Ising model and the Ising spin variables sit
 on the lattice dual to $T_2$.  

 It follows from Eq.\ \rf{7y} that the constant
 function is an eigenfunction of $K_\beta$ and $\Lambda _0 (\beta)$ is the
 corresponding eigenvalue.  Since all matrix elements of $K_\beta$ are
 positive and all entries in the eigenvector corresponding to $\Lambda_0$
 have the same sign, it follows from the Perron--Frobenius theorem that 
 $\Lambda_0$ is simple and it is the largest eigenvalue of $K_\beta$.  We
 conclude \cite{jonsson} 
 that the free energy per site in the present model is the same
 as the free energy per site in the two-dimensional Ising model.  In
 particular, the two models have the same critical point and the same
 specific heat.  However, the correlations are different, and we proceed
 to the calculation of the next eigenvalue of $K_\beta$.

 Let us introduce the notation
 \beq{8}
 \br P|Q\kt \equiv  2M^2 -2|P\triangle Q|,
 \eeq
 so $K_\beta (P,Q)=\exp \left( \beta \br P|Q\kt\right)$.
 We find it convenient to 
 think of $\br P|Q\kt $ as the `inner product'
   between the loops
 $P$ and $Q$ even though the set of loops is of course not a linear
 space.
 We shall refer to $\br P|Q\kt $ as the {\em invariant product} on $\Pi$.

 If $P$ is a loop we let $\bar{P}$ denote its complement, i.e.\ the loop
 made up of exactly those links in $T_2$ 
 that $P$ does not contain.  We let $0$
 denote the empty loop and $U$ the loop that contains all links in the
 lattice.  Then $\bar{0}=U$ and $\bar{P}=P\triangle U$.  
 The  invariant product is clearly symmetric in
 its two arguments and it has the following properties:
 \bea
 \br P|P\kt & = &2M^2\\
 \br P|0\kt & = & 2M^2-2|P|\\
 \br P|\bar{Q}\kt & = & -\br P|Q\kt.
 \eea
 One can easily construct a finite dimensional inner product space
 which contains all loops in a natural fashion such
 that the
 inner product of loops coincides with the invariant product and the
 loops lie on a sphere of radius $\sqrt{2}M$.    

 The invariant product is invariant under a nonabelian
 group $\cG$ of motions in loop-space generated by 
 \begin{itemize}
 \item
 the translations of loops
 \beq{10}
 T_a: P\mapsto P+a,
 \eeq
 $a\in T_2$,
 \item `generalized antipodal maps'
 \beq{11}
 A_Q :P\mapsto P\triangle Q,
 \eeq
 $Q\in\Pi$, 
 \item reflections and rotations in $T_2$.
 \end{itemize}
 Invariance means that
 $\br gP|gQ\kt =\br P|Q\kt$ for any $g\in\cG$. 
    As explained above we can 
 view the set of loops as a sphere and we are free to regard the
 empty loop as the north-pole and the complete 
 loop $U$ at the south-pole.  Loops
 increase in size as one moves from north to south.  The
 mapping $P\mapsto \bar{P}$ is in this picture
 the usual antipodal map and the mappings
 $A_Q$ all have the property $A_Q^2=I$ where $I$ is the
 identity map.  

 The group $\cG$ acts on the functions in $\cH$ in a natural
 way:
 \beq{11x}
 g\psi (P)=\psi (g^{-1}P).
 \eeq
 Let $O_n$ denote the linear operator on $\cH$ with matrix elements
 \beq{12}
 O_n(P,Q)=\br P|Q\kt ^n.
 \eeq
 Then $O_n$ is invariant under the action of $\cG$ on $\cH$, i.e.
 \beq{13}
 gO_ng^{-1}=O_n
 \eeq
 and the operators $O_n$ commute with each other.  Since they are symmetric 
 it follows that they
 have common eigenvectors and the same applies of course to the transfer
 matrix
 \beq{14}
 K_\beta =\sum_{n=0}^\infty {\beta ^n\over n!}\, O_n.
 \eeq
 This proves that the eigenvectors of $K_\beta$ are independent of
 $\beta$.

 In view of the invariance of the transfer matrix under $\cG$ it is
 natural to look for eigenvectors which depend on $P$ only via the
 invariant product.  Let us define
 \beq{15}
 \psi_Q (P)=\br P|Q\kt .
 \eeq
 We claim that
 \beq{16}
 K_\beta \psi _Q=\Lambda _1(\beta )\psi _Q
 \eeq
 for any $Q\in\Pi$, where 
 \beq{17}
 \Lambda _1(\beta) ={1\over 2M^2}\sum _P e^{\beta \br P|0\kt}\br P|0\kt ,
 \eeq
 so the eigenvalue ratio $\Lambda _1\Lambda _0^{-1}$ is minus the
 internal energy per link of the Ising model on $T_2$ \cite{jonsson}.

 In order to prove Eqs.\ 
 \rf{16} and \rf{17} it is convenient to introduce a function $\eta
 _{\ell}$ on $\Pi$, where $\ell$ is a link in the lattice $T_2$, defined as
 \beq{18}
 \eta _{\ell}(P)=\left\{\begin{array}{rr}-1~~~{\mbox{\rm if}}~~l\in P,\\
                                    1~~~{\mbox{\rm if}}~~l\notin P. \\ 
 \end{array}\right. 
 \eeq
 Note that
 \beq{18a}
 \br P| Q\kt=\sum _{\ell}\eta_{\ell}(P\triangle Q),
 \eeq
 where the sum runs over all links in the lattice $T_2$.
 Consider the function
 \beq{19a}
 \Phi_{Q,l}:P\mapsto \eta_{\ell} (P\triangle Q)
 \eeq
 on $\Pi$.  If $\ell\notin Q$ then $\ell\in P\triangle Q$ if and only if $\ell\in P$.
 Similarly, if $\ell\in Q$ then $\ell\in P\triangle Q$ if and only if $\ell\notin P$.  
 We conclude that
 \beq{20a}
 \eta_{\ell} (P\triangle Q)=\eta_{\ell}(P)\eta_{\ell}(Q).
 \eeq
 It follows that
 \bea
 \sum_Pe^{\beta\br 0|P\kt }\eta_{\ell}(P\triangle Q) & = &
\sum_Pe^{\beta\br 0|P\kt}\eta_{\ell}(P)\,
\eta_{\ell}(Q)\nonumber\\
 &=& {1\over 2M^2} \sum_Pe^{\beta\br 0|P\kt}\br 0|P\kt \,
\eta_\ell (Q).
 \eea
 The last equality is obtained by 
 observing that the sum 
\beq{66}
\sum_Pe^{\beta\br 0|P\kt}\eta_{\ell}(P)
\eeq
is independent of $\ell$ due to the translational and rotational invariance of 
 the invariant product and using Eq.\ \rf{18a}.
 This proves Eqs.\ \rf{16} and \rf{17}
 since it suffices to verify
 \beq{22a}
 \sum_PK_\beta (P',P)\br P|Q\kt =\Lambda_1 \br P'|Q \kt
 \eeq
 for $P'=0$ due to the invariance of $K_\beta$ under $\cG$.
In fact we have shown that the functions $\Phi_{Q,\ell}$
are all eigenfunctions of $K_\beta$ with the eigenvalue $\Lambda_1$.

 We now use the elementary functions $\Phi_{Q,\ell }$ to 
 construct the higher eigenfunctions of $K_\beta$.  Let $\ell_1$ and $\ell_2$ be 
 two distinct links.  Define
 \beq{23}
 \Phi_{Q_1Q_2,\ell_1 \ell_2}(P)=\Phi_{Q_1,\ell_1}(P)\Phi_{Q_2,\ell_2}(P).
 \eeq
 Then, by Eq.\ \rf{20a}, 
 \beq{24}
 \left(K_\beta \Phi_{Q_1Q_2,\ell_1\ell_2}\right)(0)=\Lambda_2(\ell_1,\ell_2)
 \Phi_{Q_1Q_2,\ell_1\ell_2}(0),
 \eeq
 i.e. $\Phi_{Q_1Q_2,\ell_1\ell_2}$ is an eigenfunction of $K_\beta$ with eigenvalue
 \beq{25}
 \Lambda_2(\ell_1,\ell_2)=\sum_Pe^{\beta\br 0|P\kt }\eta_{\ell_1}(P)\eta_{\ell_2}(P).
 \eeq
 Note that $\Lambda_2(\ell_1,\ell_2)$ depends on the links 
 $\ell_1$ and $\ell_2$ but only on their relative orientation 
 and the distance between them.  Of course the eigenvalue also
depends on $\beta$ but we suppress this from our notation.

 Similarly, if the links $\ell_1,\ell_2,\ldots ,\ell_n$ are all distinct, then
 \beq{26}
 \Phi_{Q_1\ldots Q_n, \ell_1 \ldots \ell_n}(P)=\prod_{i=1}^n\Phi_{Q_i,\ell_i}(P)
 \eeq
 is an eigenfunction of $K_\beta$ with the eigenvalue
 \beq{27}
 \Lambda_n(\ell_1,\ldots ,\ell_n)=\sum_Pe^{\beta\br 0|P\kt }
 \prod_{i=1}^n \eta_{\ell_i}(P).
 \eeq
 The eigenvalue is symmetric under permutations 
of the $\ell_i$'s and 
 simultaneous lattice rotations or translations of the links. 

 If $\ell$ is a link in the lattice $T_2$ let $\sigma_{\ell}^1$ and $\sigma_{\ell}^2$ be the
 two Ising spin variables on the dual sites adjacent to the link $\ell$. 
 Then $\eta_{\ell}(P)=\sigma_{\ell}^1(P)\sigma_{\ell}^2(P)$ where $\sigma_{\ell}^i(P)$ is the value
 taken by $\sigma_{\ell}^i$ in the spin configuration corresponding to the 
loop $P$.
 The expectation value 
 for an Ising model on $T_2$ is given by  
 \beq{20}
 \br (\,\cdot \,)\kt =\Lambda _0^{-1}\sum _P e^{\beta\br P|0\kt }(\,\cdot\,)
 \eeq
 so 
 \beq{28}
 \Lambda_n(\ell_1,\ldots ,\ell_n)=\Lambda_0 \, \br\prod_{i=1}^n
 \sigma_{\ell_i}^1\sigma_{\ell_i}^2\kt.
 \eeq
 The correlation inequalities
 \beq{29}
 \br\prod_{i=1}^n
 \sigma_{\ell_i}^1\sigma_{\ell_i}^2\kt \geq \br\prod_{i=1}^k
 \sigma_{\ell_i}^1\sigma_{\ell_i}^2\kt \br\prod_{i=k+1}^n
 \sigma_{\ell_i}^1\sigma_{\ell_i}^2\kt
 \eeq
 now imply
 \beq{30}
 \Lambda_n(\ell_1,\ldots ,\ell_n) \Lambda _0 \geq 
 \Lambda_k(\ell_1,\ldots ,\ell_k)
\Lambda_{n-k}(\ell_{k+1},\ldots ,\ell_n).
 \eeq
We expect that
 \beq{31}
 \Lambda_n(\ell_1,\ldots ,\ell_n)<
 \Lambda_1 ,
 \eeq
 for any $n>1$ but do not have a general proof of this inequality.
It can be checked in special cases using the explicit form of the
two spin correlation function 
\cite{wu} and the cluster property of correlations.

 \section{Multiplicities}
 As remarked above, the largest eigenvalue $\Lambda_0$ is simple by the 
 Perron--Frobenius theorem.  Let us consider the first nontrivial eigenvalue
 $\Lambda_1$ with eigenfunctions $\Phi_{Q,\ell}$. Since $\Phi_{Q,\ell}$ is a 
 multiple of $\Phi_{0,\ell}$ by the constant $\eta_{\ell}(Q)$ it suffices to 
 consider the functions $\Phi_{0,\ell}=\eta_{\ell}$.  
We claim that these functions are 
 linearly independent so the multiplicity of $\Lambda _1$ is at least $2M^2$.

 Suppose there are constants $c_{\ell}$ such that 
 \beq{32}
 \sum_{\ell}c_{\ell}\, \eta_{\ell}(P)=0
 \eeq
 for any $P\in\Pi$.  We adopt the convention that sums on $\ell$ run over all 
 links in $T_2$ unless otherwise specified.  Then
 \beq{33}
 -\sum_{\ell\in P}c_{\ell}+\sum_{\ell\notin P}c_{\ell}=0.
 \eeq
 By taking $P=0$ we also obtain
 \beq{34}
 \sum_{\ell}c_{\ell}=0.
 \eeq
 It follows that
 \beq{35}
 \sum_{\ell\in P}c_\ell =0 
 \eeq
 for any $P\in\Pi$.  Let now $x$ and $y$ be two 
different lattice points in $T_2$ and 
 suppose $P$ is made up of two simple nonintersecting curves $\gamma_1$ and 
 $\gamma_2$ 
 from $x$ to $y$.  Then clearly
 \beq{36}
 \sum_{\ell\in \gamma_1}c_{\ell}=-\sum_{\ell\in\gamma_2} c_{\ell}.
 \eeq
 Let $\gamma_3$ be one more simple curve from $x$ to $y$ 
which also avoids
 $\gamma_1$ and $\gamma_2$.  Then we can join $\gamma_3$ with either
 $\gamma_1$ or $\gamma_2$ to make a closed curve (i.e.\ a loop) in $\Pi$.
 We conclude that
 \beq{37}
 \sum_{\ell\in \gamma_1}c_{\ell}=\sum_{\ell
\in \gamma_3}c_{\ell}=\sum_{\ell\in \gamma_2}c_{\ell}
 \eeq
 which implies that 
 \beq{38}
 \sum_{\ell\in\gamma_i}c_{\ell}=0
 \eeq
 for $i=1,2,3$.  Now take $x$ and $y$ to be nearest neighbours and $\gamma _1$ 
 the curve that joins them by one link $\ell_1$.  It follows that 
$c_{\ell_1}=0$
 and hence $c_{\ell}=0$ for all $\ell$ since $x$ and $y$ are arbitrary.

We could have obtained the above result more easily by noting
that the functions $\eta_\ell$ are mutually orthogonal in the natural
inner product on $\cH$ which we denote by $(\,\cdot\, ,\,\cdot\, )$.  
This follows from the fact that $\eta_{\ell_1}
\eta_{\ell_2}$ with $\ell_1\neq\ell_2$ is an eigenfunction of $K_\beta$
with an eigenvalue $\Lambda_2(\ell_1,\ell_2)\neq \Lambda_0$ so 
$\eta_{\ell_1} \eta_{\ell_2}$ is orthogonal to the constant function, i.e.
\beq{39}
(1,\eta_{\ell_1}\eta_{\ell_2}) =\sum_P\eta_{\ell_1}(P)\eta_{\ell_2}(P)=0.
\eeq
Hence,
\beq{40}
(\eta_{\ell_1},\eta_{\ell_2})=\delta_{\ell_1\ell_2}\, 2^{M^2}
\eeq
since the total number of loops in $\Pi$ is equal to $2^{M^2}$.
This can also be seen by a direct calculation.

If the inequality \rf{31} is valid then the multiplicity of $\Lambda _1$ 
is exactly $2M^2$.  This follows from the fact that the functions
$\Phi_{Q_1\ldots Q_n,\ell_1 \ldots \ell_n}$ span
$\cH$.   We now prove this spanning property.  

Let $\ell_1,\ell_2,\ell_3,\ell_4$
be four different links which contain the same vertex.  We shall call such a 
collection of links a {\it star}.   In this case
\beq{41}
\eta_{\ell_1}(P)\eta_{\ell_2}(P)\eta_{\ell_3}(P)\eta_{\ell_4}(P)=1
\eeq
for any loop $P$ since $P$ contains $0$, $2$ or $4$ of the links $\ell_1,
\ldots ,\ell_4$.  If $\ell_1,\ell_2,\ell_3,\ell_4$ are a star 
and $n\geq 4$ it follows that
\beq{42}
\Phi_{Q_1\ldots Q_n ,\ell_1\ldots \ell_n}=
\pm \Phi_{Q_5\ldots Q_n ,\ell_5\ldots \ell_n}.
\eeq
Let $\cL =\{ \ell_1,\ldots ,\ell_n\}$ be a collection of links in $T_2$
and define
\beq{43}
E_\cL (P)=\prod_{\ell\in\cL}\eta_\ell (P).
\eeq
Clearly $E_\cL =\pm\Phi_{Q_1\ldots Q_n,\ell_1\ldots \ell_n}$ so in order 
to prove the spanning property it suffices 
to identify $2^{M^2}$ mutually orthogonal functions of the form $E_\cL$.
We adopt the convention $E_{\emptyset}=1$.

If $\cL $ and $\cL '$ are two collections of links we say that they are 
{\em equivalent} if the symmetric difference $\cL \triangle \cL '$ is
a star or the symmetric difference of two or more stars.  
This defines an equivalence relation on the 
set of all collections of links.    If $\cL$ is a collection of links 
let $[\cL ]$ denote the equivalence class of $\cL$.  It is easy to see that
the total number of elements in each equivalence class is $2^{M^2}$ 
and 
the number of different equivalence classes is also $2^{M^2}$ since 
the total number of collections of links from $T_2$ is $4^{M^2}$.  
If we choose
one $\cL _i$ from each equivalence class $[\cL _i]$, $i=1,\ldots ,2^{M^2}$, 
then
\beq{44}
(E_{\cL _i} ,E_{\cL _j})=\sum_P\prod_{\ell\in \cL _{ij}}\eta_\ell (P),
\eeq
where $\cL _{ij} $ 
is a maximal subset of $\cL _i\triangle\cL _j$ which contains 
no star.  Hence, for $i\neq j$,
\beq{45}
(E_{\cL _i},E_{\cL _j})=(1,E_{\cL _{ij}})=0
\eeq
since $\cL _{ij}$ is nonempty for $i\neq j$.  This proves that $\{ 
E_{\cL _i}\} $ is a family of mutually orthogonal functions.

\section{Discussion}
Using the results derived in the previous section it is quite easy to
calculate correlation function, i.e.\ functions of the form
\beq{46x}
G_\beta^{(N)}(P,Q)=\sum_{\pa M=P\cup Q}e^{-\beta S_1(M)},
\eeq
where the loops $P$ and $Q$ lie in two intermediate constant time planes 
separated by $N$ lattice spacings.  In Eq.\ \rf{46x} we 
sum over all surfaces whose intersection
with the two intermediate planes are
$P$ and $Q$ and  $S_1(M)$ is 
a modified action functional, counting only edges that are orthogonal to 
the time direction.  We have
\beq{47}
G_\beta^{(N)}(P,Q)=(\delta_P,K_\beta^N\delta_Q),
\eeq
where $\delta_P$ and $\delta_Q$ are delta functions 
in $\cH$.  Up to normalization we can
interpret $G_\beta^{(N)}(P,Q)$ as the probability of having the loop 
$P$ at time $N$ given that we have $Q$ at time $0$.  
An easy calculation gives
\beq{48}
G_\beta^{(N)}(P,Q)=2^{-M^2}\sum_{i=1}^{2^{M^2}} E_{\cL _i} 
(P\triangle Q)\Lambda_{\cL _i}^N,
\eeq
where the sum runs over the orthogonal family of functions constructed 
in the last section and $\Lambda_{\cL _i}$
is the eigenvalue of $K_\beta$ corresponding to $E_{\cL _i}$.

Unfortunately it is not clear how to extend the analysis of the present 
paper to include the curvature terms in the action of the full model.
The original transfer matrix \rf{4} is not invariant under the group $\cG$, 
only translations and rotations remain symmetries.  The constant 
function is not an eigenfunction any more.   The only result that 
survives is the fact that the largest eigenvalue is in this case also
simple by the Perron--Frobenius theorem.

However, we have managed to solve exactly a three dimensional lattice model
that describes the diffusion of loops.  As a surface model
it is not isotropic but can be viewed as a stack of two-dimensional 
Ising systems with
an intereaction that is sufficiently simple for us to solve the model exactly.

\bigskip

\noindent
{\bf Acknowledgement.} T.~J. is indebted to NRCPS ``Demokritos'' and the 
CERN Theory Division for hospitality.

\end{document}